# A low-phase-noise 18 GHz Kerr frequency microcomb phase-locked over 65 THz

S.-W. Huang[1,2,*], J. Yang[1,2], H. Zhou[1], M. Yu[3], D.-L. Kwong[3], and C. W. Wong[1,2,*]

[1] Mesoscopic Optics and Quantum Electronics Laboratory, University of California, Los Angeles, CA 90095

[2] Optical Nanostructures Laboratory, Center for Integrated Science and Engineering, Solid-State Science and Engineering, and Mechanical Engineering, Columbia University, New York, NY 10027

[3] Institute of Microelectronics, Singapore, Singapore 117685

*Author e-mail address: swhuang@seas.ucla.edu, cheewei.wong@ucla.edu

Laser frequency combs are coherent light sources that simultaneously provide pristine frequency spacings for precision metrology and the fundamental basis for ultrafast and attosecond sciences. Recently, nonlinear parametric conversion in high-$Q$ microresonators has been suggested as an alternative platform for optical frequency combs, though almost all in 100 GHz frequencies or more. Here we report a low-phase-noise on-chip Kerr frequency comb with mode spacing compatible with high-speed silicon optoelectronics. The waveguide cross-section of the silicon nitride spiral resonator is designed to possess small and flattened group velocity dispersion, so that the Kerr frequency comb contains a record-high number of 3,600 phase-locked comb lines. We study the single-sideband phase noise as well as the long-term frequency stability and report the lowest phase noise floor achieved to date with -130 dBc/Hz at 1 MHz offset for the 18 GHz Kerr comb oscillator, along with feedback stabilization to achieve frequency Allan deviations of $7\times10^{-11}$ in 1 s. The reported system is a promising compact platform for achieving self-referenced Kerr frequency combs and also for high-capacity coherent communication architectures.

Optical frequency combs, since their inception more than a decade ago [1], has led to breakthroughs in precision spectroscopy [2,3], frequency metrology [4,5], and astrophysical spectrography [6,7]. They are also promising platforms for optical communication [8,9], stable microwave signal generation [10], and arbitrary optical waveform generation [11]. The current benchmark laser systems for optical frequency combs are self-referenced femtosecond mode-locked lasers [12]. However, continuous-wave (cw) pumped microresonators recently emerge as promising alternative platforms for optical frequency comb generation [13]. Frequency combs here are generated by modulation instability and four wave mixing, facilitated by the high quality factors and small mode volumes of these microresonators. Microresonator-based optical frequency combs, or Kerr frequency combs, are unique in their compact footprints and offer the potential for monolithic electronic and feedback integration, thereby expanding the already remarkable applications of frequency combs. To this end, microresonator-based optical frequency combs with comb spacings of 10 to 40 GHz, compatible with high-speed optoelectronics, have recently been examined in whispering gallery mode (WGM) structures [14–20] and planar ring geometries [9,21]. Planar ring cavities are particularly attractive since: 1) the resonator and the coupling waveguide can be monolithically integrated, reducing the sensitivity to the environmental perturbation; 2) the resonator only supports a few discrete transverse modes, increasing the robustness of coupling into the designed resonator mode family; and 3) the cavity dispersion and the comb spacing can be engineered separately, offering the flexibility to tailor the cavity dispersion for efficient and broadband comb generation.

Here we report a low-phase-noise Kerr frequency comb generated from a silicon nitride spiral resonator. With the small and flattened group velocity dispersion, the 18 GHz Kerr frequency comb spans nearly half an octave and contains a record-high number of comb lines at more than 3,600. Spectral modulation induced by mode interactions is also evidently observed. A single bandwidth-limited RF beat note is observed and the single-sideband (SSB) phase noise analysis reveals the lowest phase noise floor achieved to date in free-running Kerr frequency combs, -130 dBc/Hz at 1 MHz offset for the 18 GHz carrier. The long-term frequency stability is characterized and the measured free-running Allan deviation is $2\times10^{-8}$ in 1 s, consistent with the frequency fluctuations caused by the pump wavelength drift. Feedback stabilization further improves the frequency stability to $7\times10^{-11}$ in 1 s.

Figure 1a shows an optical micrograph of the silicon nitride spiral resonator and the cavity dispersion simulated with full-vector finite-element mode solver. The microresonator is fabricated with CMOS-compatible processes for the low-pressure chemical vapor deposition of the nitride and it is annealed at a temperature of 1200$^\circ$C to reduce the N-H overtone absorption. The spiral design ensures the relatively large resonator fits into a tight field-of-view to avoid stitching and discretization errors during the photomask generation [21], which can lead to higher cavity losses. Bends in the resonator have diameters greater than 160 μm to minimize the bending-induced dispersion. The waveguide cross-section is designed to be 2 μm × 0.75 μm so that not only the group velocity dispersion (GVD) but also the third order dispersion (TOD) is small in this microresonator. The small and flattened GVD is critical for broadband comb generation [22]. Figure 1b shows the pump mode is critically coupled with a loaded quality factor approaching 660,000 (intrinsic quality factor at 1,300,000). A tunable external-cavity diode laser (ECDL) is amplified by an L-band erbium doped fiber amplifier (EDFA) to 2W and then coupled to the microresonator with a single facet coupling loss of 3 dB, resulting in a coupled pump power 5 times higher than the threshold pump power. A 1583-nm longpass filter is used to remove the amplified spontaneous emission noise from the EDFA. Both the pump power and the microresonator chip's temperature are actively stabilized such that the fluctuation of the on-chip pump power is less than $10^{-3}$. A 3-paddle fiber polarization controller and a polarization beam splitter cube are used to ensure the proper coupling of TE polarization into the microresonator. To obtain the Kerr frequency comb, the pump wavelength is first tuned into the resonance from the high frequency side at a step of 1 pm (~118 MHz) until a broadband comb is observed on the optical spectrum analyzer. Importantly, it is then necessary to switch to fine control of the pump wavelength at a step of <5 MHz in order to drive the comb from a noisy state to a phase-locked state. At the output, 5-nm WDM filters are used to notch the pump and a dispersion compensating fiber jumper is used to properly cancel the dispersion introduced by the WDM filters. An example of the Kerr frequency comb is shown in Figure 1c, spanning nearly half an octave (65 THz, defined as 60 dB below the maximum comb line power) and covering multiple telecommunication bands (E, S, C, L and U bands) with the comb spacing of 17.986 GHz. Of note, the generated Kerr frequency comb contains more than 3,600 comb lines, the record large number of Kerr comb lines made possible by the small and flattened GVD.

Two TE modes with different free spectral ranges (18 and 17.4 GHz) are supported in the spiral resonator and their resonance wavelengths periodically get close to each other with a period of ≈ 4 nm. Figure 2a (top) plots the resonance wavelength offsets of the second-order mode family with respect to the fundamental mode family. The zero crossings (red horizontal line) represent the wavelengths where the resonances of the two mode families are supposed to be degenerate. However, the degeneracy is lifted due to the mode interaction, as evidenced by the openings in the resonance wavelength offsets around the zero crossings. Such anti-crossing phenomena leads to the local disruption of dispersion and modifies the phase matching condition of the comb generation process [23–26]:

$$\Delta k(\omega_{FSR}\mu) = \beta_2 \omega_{FSR}^2 \mu^2 + \kappa(\omega_{FSR}\mu) + \gamma P_{int} - \delta$$

where $\beta_2$ is the GVD, $\omega_{FSR}$ the free spectral range, $\mu$ the mode number, $\kappa(\omega_{FSR}\mu)$ the periodic local dispersion disruption by the mode interaction, $\gamma$ the nonlinear coefficient, $P_{int}$ the intracavity pump power, and $\delta$ the pump wavelength detuning. While the local dispersion disruption is 2 to 3 GHz, the GVD is only 20 fs$^2$/mm and it takes ≈ 1,000 modes before the GVD induced phase mismatch becomes comparable to that induced by the mode interaction. Thus the phase matching condition around the pump should be dominated by the mode interaction, as evidently shown in Figure 2a where the correlation between the zero crossings (top) and the local maxima of the Kerr frequency comb (bottom) is observed.

Figure 2b shows the RF amplitude noise spectra of the Kerr frequency comb [18,27]. When the primary comb line spacing is incommensurate with the fundamental comb spacing, multiple RF peaks will occur due to the beating between different comb families (Figure 2b inset). Next, with fine control of the pump wavelength, the offset between different comb families can be made zero such that the RF amplitude noise spectrum shows no excess noise (Figure 2b). To characterize the RF beat note of the 18 GHz Kerr frequency comb, a high-speed photodetector is used to demodulate the frequency comb at 17.986 GHz, and an 18.056 GHz local oscillator is used to downmix the electronic signal to the baseband for analysis. Figure 2c plots the RF spectra of the beat notes from three different filtered spectral regions of the comb (black curve: whole spectrum excluding the pump; blue curve: 1529 to 1538 nm; red curve: 1555 to 1564 nm). The pedestal below 500 kHz offset frequency comes from the 18.056 GHz local oscillator. All three measurements show bandwidth-limited beat notes at 17.986 GHz, characteristic of an equidistant Kerr frequency comb as those of a non-equidistant comb will either reside at distinct frequencies or show different linewidths and phase noise characteristics (Figure 3) [21]. A second-harmonic-generation optical intensity autocorrelation is implemented to characterize the temporal structure of the phase-locked Kerr frequency comb and the result is shown in Figure 2d. The trace has a contrast of ≈ 2, characteristic of a comb without a clean circulating high-peak power pulse. Recently, a phase-locked Kerr frequency comb without mode-locking is also observed in a WGM microresonator [20] and the self-injection locking is shown to be the underlying mechanism for driving the Kerr frequency comb into a phase-locked state [19,20]. Compared to smaller microresonators where mode-locking are demonstrated [25,28,29], self-injection locking plays a more important role in low-repetition-rate Kerr frequency combs because the more frequent local dispersion disruptions impede the mode-locking from occurring [30].

Figure 3 shows the SSB phase noise spectra of the RF beat notes. To probe the possibility of degraded phase noise for different spectral slices of the comb due to the complicated nonlinear comb generation process [18], here again we measure the phase noises at three different spectral regions (A, blue curve: 1529 to 1538 nm; B, red curve: 1555 to 1564 nm; C, black curve: whole spectrum excluding the pump). Compared to the comb lines in the region A, the comb lines in the region B are characterized by their higher optical power and better amplitude uniformity. However, the phase noise results show that the mechanism for phase noises at different parts of the Kerr frequency comb is identical and the minutely better phase noise floor of the region B and C is a direct consequence of the higher comb power. The olive line shows the phase noise of the local oscillator used for downmixing the RF beat note of Kerr frequency comb and it is worth mentioning that the phase noise of the comb beat note becomes comparable or better than that of the commercially available high performance microwave local oscillators for offset frequencies higher than 20 kHz. The record low phase noise floor of -130 dBc/Hz, as well as the record large number of Kerr comb lines, can be well-suited for further improving the high-capacity coherent data transmission with advanced phase modulation techniques [9]. For offset frequency below 10 kHz, the phase noise is dominated by frequency flicker (30 dB/decade) which can be accounted by noise induced from the wavelength drift of the pump laser.

Figure 4 shows the Allan deviation of the free-running (open squares) and the stabilized (closed squares) Kerr frequency comb spacing. The free-running Allan deviation is measured at $2\times10^{-8}$ in 1 s, increase to $1\times10^{-7}$ in 10 s, and then gradually decrease to $5\times10^{-8}$ in 100 s. As the detuning changes the intracavity power, the comb spacing linearly shifts by 57 Hz per MHz of pump wavelength detuning in our microresonator (Figure 4 inset). On the other hand, the instability of the employed pump laser wavelength is characterized by measuring the heterodyne beat note between the pump laser and a tooth of a fiber frequency comb. The estimated comb spacing instability caused by the pump wavelength drift is also plotted in Figure 4 (red diamond) and it shows the pump wavelength drift is the main noise source limiting the long term stability of the Kerr frequency comb spacing. Finally, a proportional-integral feedback stabilization of the comb spacing by controlling the pump wavelength is implemented to improve the long-term stability by more than two orders of magnitude, reaching $7\times10^{-11}/\tau^{0.84}$, comparable to that of a commercially available high performance microwave oscillator.

In summary, we report a low-phase-noise Kerr frequency comb generated from a silicon nitride spiral resonator. The 18 GHz Kerr frequency comb spans nearly half an octave and contains a record-high number of comb lines at ~3,600. We study the SSB phase noise and report the lowest phase noise floor achieved to date in Kerr frequency combs, -130 dBc/Hz at 1 MHz offset for 18 GHz carrier. Limited by the wavelength drift of the employed pump laser, the free-running Allan deviation is measured at $2\times10^{-8}$ in 1 s and it is improved to $7\times10^{-11}$ at 1 s by a feedback control of the pump wavelength. With half-octave-spanning bandwidth, record large number of Kerr comb lines, and record low phase noise floor, the reported system is a promising compact platform not only for achieving self-referenced Kerr frequency combs but also for improving high-capacity coherent telecommunication systems. Although ultrashort pulses are not generated directly from this microresonator, the high-coherence phase locking property lends itself to pulse shaping technique for harvesting the temporal features of the Kerr frequency comb [11].

**Methods**

**Device fabrication**: First a 3 μm thick $SiO_2$ layer was deposited via plasma-enhanced chemical vapor deposition on p-type 8" silicon wafers to serve as the under-cladding oxide. Then low-pressure chemical vapor deposition (LPCVD) was used to deposit a 750 nm silicon nitride for the spiral resonators, with a gas mixture of $SiH_2Cl_2$ and $NH_3$. The resulting $Si_3N_4$ layer was patterned by optimized 248 nm deep-ultraviolet lithography and etched down to the buried $SiO_2$ via optimized reactive ion dry etching. The sidewalls were observed under SEM for an etch verticality of 88 degrees. The silicon nitride spiral resonators were then over-cladded with a 3 μm thick $SiO_2$ layer, deposited initially with LPCVD (500 nm) and then with plasma-enhanced chemical vapor deposition (2500 nm).

**Acknowledgements:** The authors acknowledge discussions with James F. McMillan and Yongjun Huang, and loan of the microwave signal generator and the high-speed RF frequency mixer from the Bergman group and Krishnaswamy group, respectively, at Columbia University.

**Author contributions:** S.W.H. designed and performed the experiment, analyzed the data and wrote the paper. S.W.H., J.H.Y., and C.W.W. designed the layout. M.Y. and D.L.K. performed the device nanofabrication. S.W.H., H.Z., J.H.Y., and C.W.W. contributed to discussion and revision of the manuscript.

**Additional information:** The authors declare no competing financial interests. Reprints and permission information is available online at http://www.nature.com/reprints/. Correspondence and requests for materials should be addressed to S.W.H and C.W.W.

**References:**

[1] J. Reichert, R. Holzwarth, T. Udem, and T. W. Hänsch, "Measuring the frequency of light with mode-locked lasers" *Opt. Commun.* **172**, 59 (1999).

[2] R. Holzwarth, T. Udem, T. W. Hänsch, J. C. Knight, W. J. Wadsworth, and P. S. J. Russell, "Optical Frequency Synthesizer for Precision Spectroscopy" *Phys. Rev. Lett.* **85**, 2264 (2000).

[3] A. Cingöz, D. C. Yost, T. K. Allison, A. Ruehl, M. E. Fermann, I. Hartl, and J. Ye, "Direct frequency comb spectroscopy in the extreme ultraviolet" *Nature* **482**, 68 (2012).

[4] T. Udem, R. Holzwarth, and T. W. Hänsch, "Optical frequency metrology" *Nature* **416**, 233 (2002).

[5] L.-S. Ma, Z. Bi, A. Bartels, L. Robertsson, M. Zucco, R. S. Windeler, G. Wilpers, C. Oates, L. Hollberg, and S. A. Diddams, "Optical Frequency Synthesis and Comparison with Uncertainty at the $10^{-19}$ Level" *Science* **303**, 1843 (2004).

[6] C.-H. Li, A. J. Benedick, P. Fendel, A. G. Glenday, F. X. Kartner, D. F. Phillips, D. Sasselov, A. Szentgyorgyi, and R. L. Walsworth, "A laser frequency comb that enables radial velocity measurements with a precision of $1\,\mathrm{cm\,s^{-1}}$" *Nature* **452**, 610 (2008).

[7] T. Wilken, G. L. Curto, R. A. Probst, T. Steinmetz, A. Manescau, L. Pasquini, J. I. G. Hernández, R. Rebolo, T. W. Hänsch, T. Udem, and R. Holzwarth, "A spectrograph for exoplanet observations calibrated at the centimetre-per-second level" *Nature* **485**, 611 (2012).

[8] J. S. Levy, A. Gondarenko, M. A. Foster, A. C. Turner-Foster, A. L. Gaeta, and M. Lipson, "CMOS-compatible multiple-wavelength oscillator for on-chip optical interconnects" *Nature Photon.* **4**, 37 (2010).

[9] J. Pfeifle, V. Brasch, M. Lauermann, Y. Yu, D. Wegner, T. Herr, K. Hartinger, P. Schindler, J. Li, D. Hillerkuss, R. Schmogrow, C. Weimann, R. Holzwarth, W. Freude, J. Leuthold, T. J. Kippenberg, and C. Koos, "Coherent terabit communications with microresonator Kerr frequency combs" *Nature Photon.* **8**, 375 (2014).

[10] T. M. Fortier, M. S. Kirchner, F. Quinlan, J. Taylor, J. C. Bergquist, T. Rosenband, N. Lemke, A. Ludlow, Y. Jiang, C. W. Oates, and S. A. Diddams, "Generation of ultrastable microwaves via optical frequency division" *Nature Photon.* **5**, 425 (2011).

[11] F. Ferdous, H. Miao, D. E. Leaird, K. Srinivasan, J. Wang, L. Chen, L. T. Varghese, and A. M. Weiner, "Spectral line-by-line pulse shaping of on-chip microresonator frequency combs" *Nature Photon.* **5**, 770 (2011).

[12] J. Ye and S. T. Cundiff, *Femtosecond Optical Frequency Comb: Principle, Operation and Applications* (Springer, New York, NY, 2005).

[13] T. J. Kippenberg, R. Holzwarth, and S. A. Diddams, "Microresonator-Based Optical Frequency Combs" *Science* **332**, 555 (2011).

[14] A. A. Savchenkov, A. B. Matsko, V. S. Ilchenko, I. Solomatine, D. Seidel, and L. Maleki, "Tunable Optical Frequency Comb with a Crystalline Whispering Gallery Mode Resonator" *Phys. Rev. Lett.* **101**, 093902 (2008).

[15] I. S. Grudinin, N. Yu, and L. Maleki, "Generation of optical frequency combs with a $CaF_2$ resonator" *Opt. Lett.* **34**, 878 (2009).

[16] Y. K. Chembo, D. V. Strekalov, and N. Yu, "Spectrum and Dynamics of Optical

Frequency Combs Generated with Monolithic Whispering Gallery Mode Resonators" *Phys. Rev. Lett.* **104**, 103902 (2010).

[17] W. Liang, A. A. Savchenkov, A. B. Matsko, V. S. Ilchenko, D. Seidel, and L. Maleki, "Generation of near-infrared frequency combs from a $MgF_2$ whispering gallery mode resonator" *Opt. Lett.* **36**, 2290 (2011).

[18] T. Herr, K. Hartinger, J. Riemensberger, C. Y. Wang, E. Gavartin, R. Holzwarth, M. L. Gorodetsky, and T. J. Kippenberg, "Universal formation dynamics and noise of Kerr-frequency combs in microresonators" *Nature Photon.* **6**, 480 (2012).

[19] J. Li, H. Lee, T. Chen, and K. J. Vahala, "Low-Pump-Power, Low-Phase-Noise, and Microwave to Millimeter-Wave Repetition Rate Operation in Microcombs" *Phys. Rev. Lett.* **109**, 233901 (2012).

[20] P. Del'Haye, K. Beha, S. B. Papp, and S. A. Diddams, "Self-Injection Locking and Phase-Locked States in Microresonator-Based Optical Frequency Combs" *Phys. Rev. Lett.* **112**, 043905 (2014).

[21] A. R. Johnson, Y. Okawachi, J. S. Levy, J. Cardenas, K. Saha, M. Lipson, and A. L. Gaeta, "Chip-based frequency combs with sub-100 GHz repetition rates" *Opt. Lett.* **37**, 875 (2012).

[22] L. Zhang, C. Bao, V. Singh, J. Mu, C. Yang, A. M. Agarwal, L. C. Kimerling, and J. Michel, "Generation of two-cycle pulses and octave-spanning frequency combs in a dispersion-flattened micro-resonator" *Opt. Lett.* **38**, 5122 (2013).

[23] A. A. Savchenkov, A. B. Matsko, W. Liang, V. S. Ilchenko, D. Seidel, and L. Maleki, "Kerr frequency comb generation in overmoded resonators" *Opt. Express* **20**, 27290 (2012).

[24] I. S. Grudinin, L. Baumgartel, and N. Yu, "Impact of cavity spectrum on span in microresonator frequency combs" *Opt. Express* **21**, 26929 (2013).

[25] S.-W. Huang, H. Zhou, J. Yang, J. F. McMillan, A. Matsko, M. Yu, D.-L. Kwong, L. Maleki, and C. W. Wong, "Mode-locked ultrashort pulse generation from on-chip normal dispersion microresonators" *Phys. Rev. Lett.* (accepted to appear; Feb 2015).

[26] Y. Liu, Y. Xuan, X. Xue, P.-H. Wang, S. Chen, A. J. Metcalf, J. Wang, D. E. Leaird, M. Qi, and A. M. Weiner, "Investigation of mode coupling in normal-dispersion silicon

nitride microresonators for Kerr frequency comb generation" *Optica* **1**, 137 (2014).

[27] C. Y. Wang, T. Herr, P. Del'Haye, A. Schliesser, J. Hofer, R. Holzwarth, T. W. Hänsch, N. Picqué, and T. J. Kippenberg, "Mid-infrared optical frequency combs at 2.5 μm based on crystalline microresonators" *Nature Commun.* **4**, 1345 (2013).

[28] K. Saha, Y. Okawachi, B. Shim, J. S. Levy, R. Salem, A. R. Johnson, M. A. Foster, M. R. E. Lamont, M. Lipson, and A. L. Gaeta, "Modelocking and femtosecond pulse generation in chip-based frequency combs" *Opt. Express* **21**, 1335 (2013).

[29] T. Herr, V. Brasch, J. D. Jost, C. Y. Wang, N. M. Kondratiev, M. L. Gorodetsky, and T. J. Kippenberg, "Temporal solitons in optical microresonators" *Nature Photon.* **8**, 145 (2014).

[30] T. Herr, V. Brasch, J. D. Jost, I. Mirgorodskiy, G. Lihachev, M. L. Gorodetsky, and T. J. Kippenberg, "Mode Spectrum and Temporal Soliton Formation in Optical Microresonators" *Phys. Rev. Lett.* **113**, 123901 (2014).

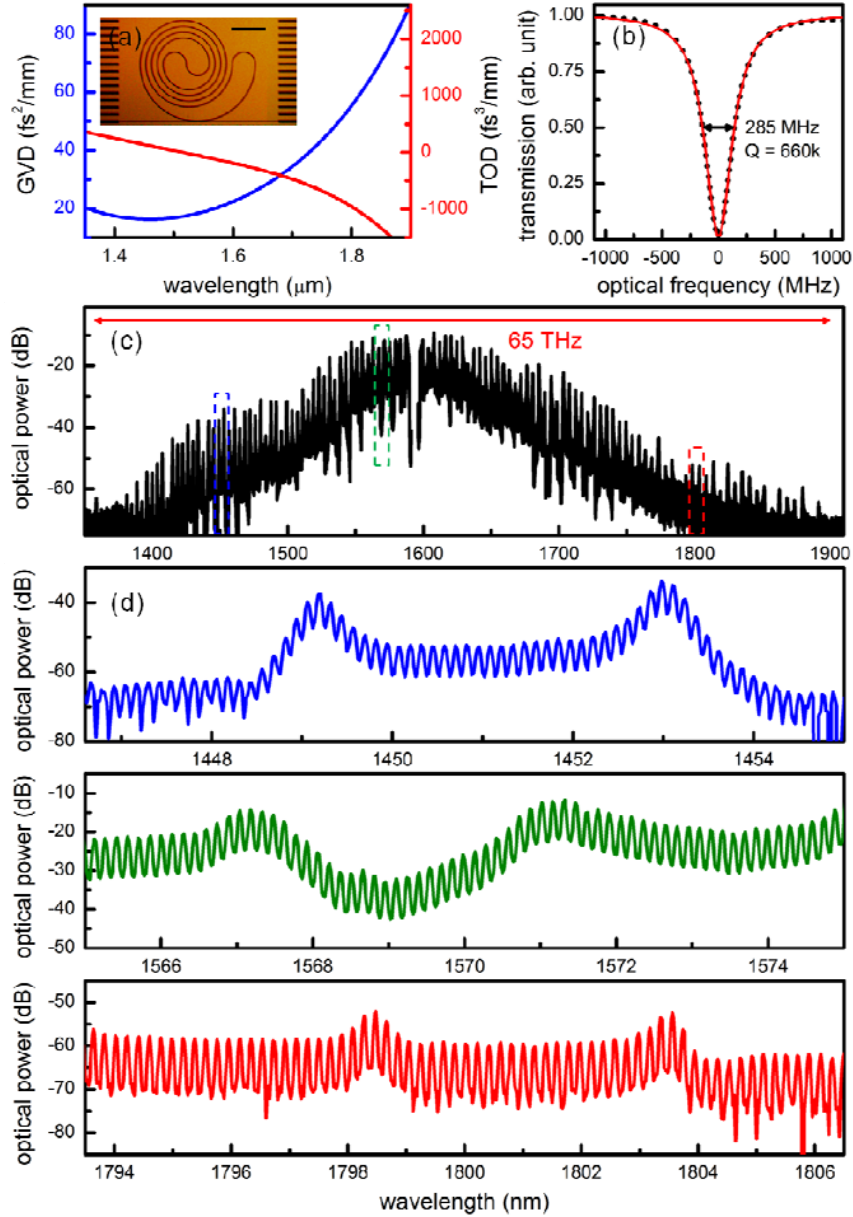

**Figure 1 | A phase-locked 18 GHz optical frequency comb spanning over 65 THz. a,** Simulated group velocity dispersion (GVD) and third order dispersion (TOD) of the ring resonator, featuring small TOD which is beneficial for broad comb generation. Inset: An optical micrograph of the spiral resonator, with a total cavity length of 8.04 mm and a mode area of 1.3 μm$^2$. Adiabatic mode converters (the dark bars on the side of the chip) are implemented to improve the coupling efficiency from the free space to the bus waveguide (the bottom straight line across the chip). Scale bar: 250 μm. **b,** Example critically-coupled resonant pump mode at 1595.692 nm, with a 285 MHz loaded cavity linewidth. Black dots are the measured data points and the red curve is the fitted Lorentzian lineshape. **c,** Example

generated Kerr frequency comb, with a broad spectrum spanning nearly half an octave at 65 THz and covering multiple telecommunication bands (E, S, C, L and U bands). **d,** Zoom-in views of the comb spectra from 1446.5 nm to 1455 nm (blue), 1565 nm to 1575 nm (green), and 1793.5 nm to 1806.5 nm (red). Even in the wings of the spectrum, native-FSR-spacing comb lines are clearly observed.

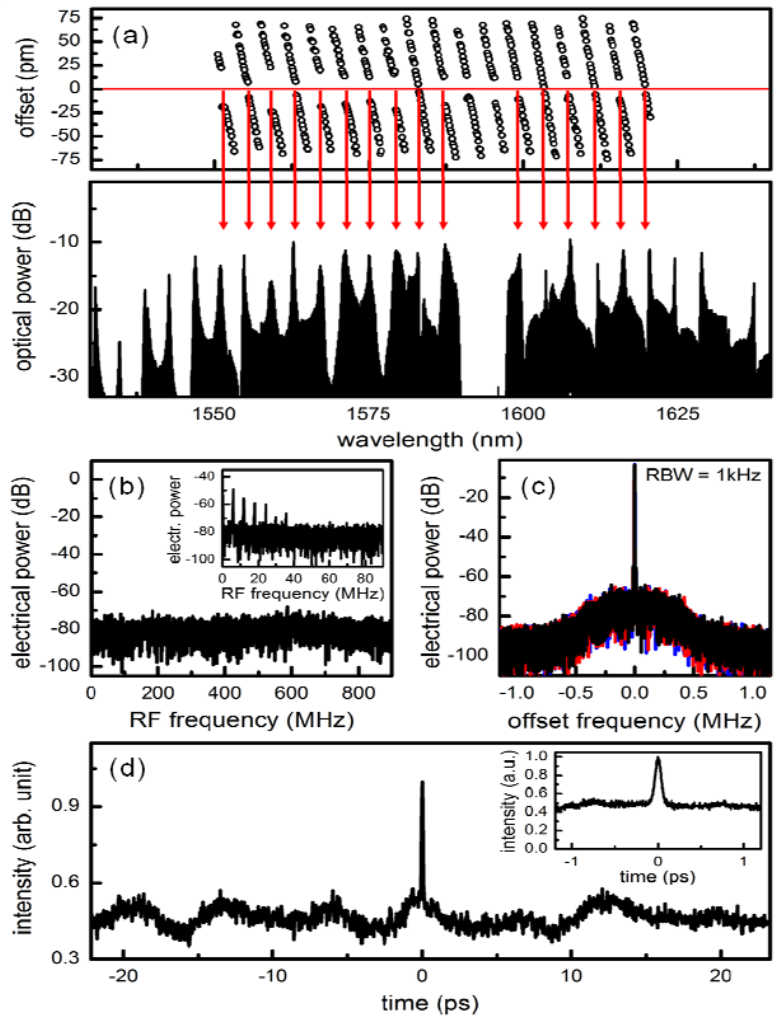

**Figure 2 | Fundamental and second-order family offsets and noise characteristics of the phase-locked 18 GHz comb. a,** Resonance frequency offsets of the second-order mode family with respect to the fundamental mode family (top) as well as the zoom-in view of the Kerr frequency comb (bottom). **b,** RF amplitude noise of the offset-free Kerr frequency comb. Inset: an example RF amplitude noise of the Kerr frequency comb showing multiple peaks due to the beating between different comb families. The comb is tuned to be offset-free by fine control of the pump wavelength. **c,** RF spectra of the beat notes from three different filtered spectral regions of the comb (black curve: whole spectrum excluding the pump; blue curve: 1529 to 1538 nm; red curve: 1555 to 1564 nm). All three measurements show bandwidth-limited beat notes at 17.986 GHz, characteristic of a phase-locked comb. The pedestal below 500 kHz offset frequency comes from the local oscillator used for downmixing the 17.986 GHz beat note (Figure 3). **d,** Optical intensity autocorrelation of the phase-locked Kerr frequency comb. The trace has a contrast of ~ 2, characteristic of a comb without a clean circulating high-peak power pulse.

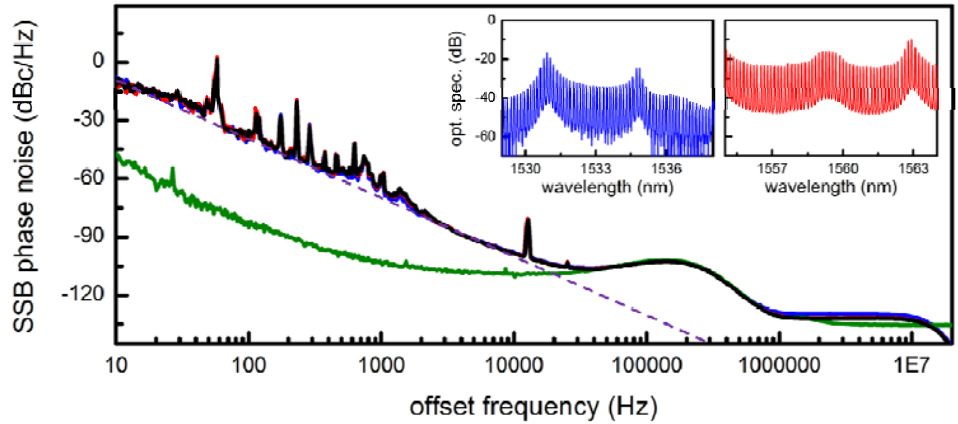

**Figure 3 | Single-sideband (SSB) phase noises of the beat notes from different spectral regions of the comb.** Three filtered spectral regions of the comb with the comb lines are shown in the inset (blue curve: 1529 to 1538 nm; red curve: 1555 to 1564 nm). The black curve shows the whole spectrum excluding the pump. All SSB phase noise spectra show a very low phase noise floor of -130 dBc/Hz at 1 MHz offset from the carrier. For offset frequency below 10 kHz, the phase noise has a roll-off of 30 dB/decade (purple dashed line). The olive curve is the SSB phase noise of the local oscillator used for downmixing the 17.986 GHz beat note.

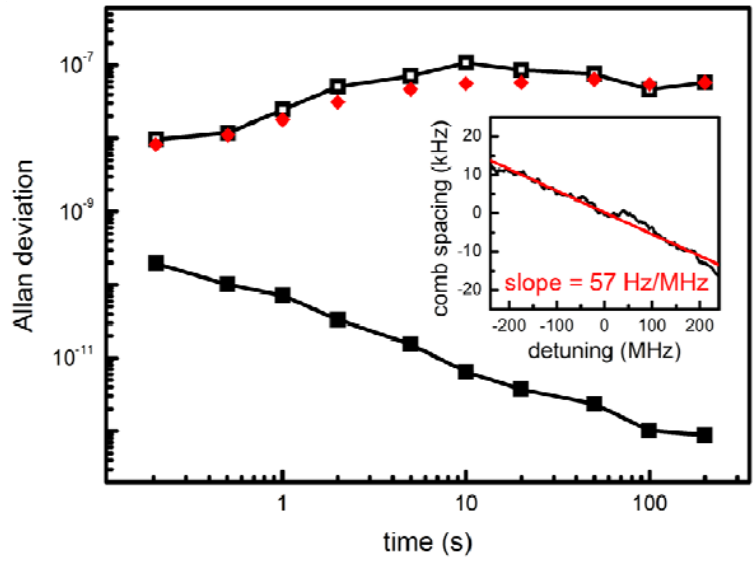

**Figure 4 | Allan deviation of the free-running (open squares) and the stabilized (closed squares) Kerr frequency comb spacing.** In free running mode, the comb spacing stability is limited by the fluctuation resulting from the pump laser wavelength drift (red diamond). Feedback stabilization is achieved by monitoring the comb spacing and controlling the pump laser wavelength to compensate the errors with a proportional-integral controller. Inset: The comb spacing as a function of the pump wavelength detuning, determined at 57 Hz/MHz in our microresonator.